\documentclass[aps,prl,twocolumn,showpacs, floatfix]{revtex4}
\usepackage{graphicx}
\usepackage{amsmath}

\begin{document}

\title{Spontaneous development of 3-D structure in sheared granular flows}
\author{James F. Lutsko}
\email{jlutsko@ulb.ac.be}
\affiliation{Center for Nonlinear Phenomena and Complex Systems\\
Universit\'{e} Libre de Bruxelles\\
Campus Plaine, CP 231, 1050 Bruxelles, Belgium}
\date{\today }

\begin{abstract}
Computer simulations of sheared granular fluids, modeled as inelastic hard
spheres, are presented which show signs of a uniquely three-dimensional
instabilty. In the stable regime, a linear velocity profile, $v_{x}=ay$,
with shear rate $a$ is established using Lees-Edwards boundary conditions.
In the unstable regime, the velocity profile aquires a dependence on the
third dimension of the form $v_{x}=ay+sin(2\pi z/L)$ in a cubic box with
sides of length $L$. An analysis of the linearized Navier-Stokes equations
shows the presence of an instability and gives a simple expression for the
critical wavevector which is quantitatively consistent with the results of
simulations and which indicates that the instability persists at low
densities.
\end{abstract}

\maketitle

One of the intriguing aspects of fluidized granular flows is that they
exhibit a rich phenomenology of clustering, segregation and pattern
formation under different flow conditions(see e.g. \cite{CampbellReview},%
\cite{GranPhysToday},\cite{GranRMP}). Understanding the mechanisms that give
rise to this behavior is one of the principle challenges in the field of
granular physics. The minimal microscopic model of a granular fluid as hard
spheres which lose energy upon collision is enough to give rise to many
effects mentioned above. For example, computer simulations using this model
of an isolated fluid shows that it is subject to both linear instabilities,
which drive the formation of vortices\cite{McNamaraYoung},\cite{Orza},\cite%
{BreyDuftyKimSantos}, and nonlinear instabilities which lead to the
formation of clusters\cite{GoldZan},\cite{BreyHCS}. Similar phenomena are
observed in both simulation and experiment of the more practically relevant
sheared granular flows, in which the flow speed varies in some direction
perpendicular to the direction of flow. Most attention has been focussed on
2-D flows due to the relative computational efficiency with which they can
be simulated and ease with which they can be studied in experiments and the
results visualized although a few experiments have probed the 3-dimensional
structure in annular geometries\cite{GranGollub}. It has long been known
that 2-D flows show the formation of both large scale structures (shear
bands) and small-scale ordering of the grains\cite{TanGold}. The analysis of
the stability of shear flow is complicated by the convective terms which
give rise to a coupling of hydrodynamic modes\cite{Lutsko_Fluctuations} for
wavevectors that sample the direction of the flow but it has been shown that
2-D unbounded shear flows are linearly stable though subject to nonlinear
instabilities that drive the pattern formation\cite{TanGold}. Bounded 2-D
shear flows, with moving walls perpendicular to the velocity gradient and
periodic boundaries in the direction of the flow, have been shown to be
linearly unstable\cite{Sasvari},\cite{AlamNott}. Dense granular flows in 3
dimensions show solid-like ordering into regular arrays\cite{GranGollub}
reminiscent of that seen in elastic sheared fluids\cite{ErpShear},\cite%
{Lutsko96}.

In this work, I present the results of computer simulations of the minimal
granular model subject to shear flow in an unbounded geometry using
Lees-Edwards boundary conditions and show that the system develops
3-dimensional structure even at \emph{low} densities where there is no sign
of the solid-like ordering. In particular, the system develops an
intermittant variation of the peculiar velocity (relative to the linear
profile) as a function of position in the direction perpendicular to both
the flow and the velocity gradient with clustering near the extrema of this
profile. A linear stability analysis of the relevant hydrodynamic equations
shows the presence of a linear instability towards perturbations with
wavevectors in the third dimension and a simple stability criterion is
derived. Interestingly, this instability appears to be very similar to one
which occurs in the stability analysis of the HCS\cite{DeltourHCS},\cite%
{ErnstBritoHCS},\cite{BreyDuftyKimSantos} but which is hidden by the
nonlinear instabilities\cite{BreyHCS}. Based on these results it appears
that 3-D sheared granular flows are linearly unstable for all densities.

\begin{figure}
\includegraphics[angle=-90,scale=0.3]{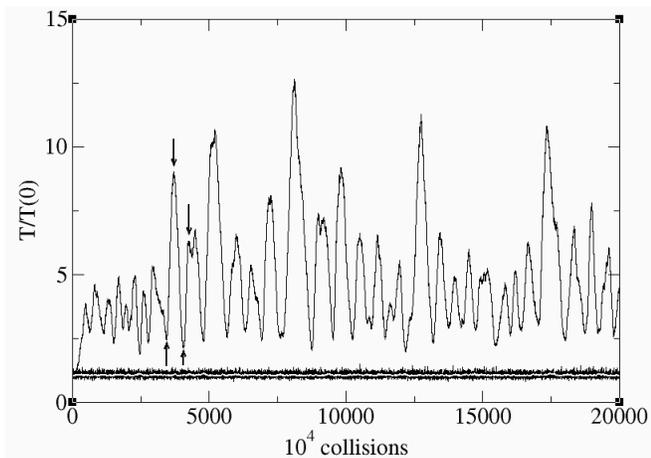}
\caption{\label{fig1} Temperature as a function of the number of collisions for $\alpha
=0.7$ with $13500$ grains, upper black line, $500$ grains, lower black line,
and $\alpha =0.9,N=13500$. white line. The arrows mark the times at which
snapshots were taken. The arrows, from left to right, mark the snapshots
(a)-(d) of the other figures.}
\end{figure}

The simulations described here of mechanically identical atoms were
performed using an event-driven algorithm in which atoms stream freely
between collisions. Two atoms with positions and velocities $\overrightarrow{%
q}_{i},\overrightarrow{v}_{i}$ and $\overrightarrow{q}_{j},\overrightarrow{v}%
_{j}$ collide when their separation $q_{ij}\equiv \left| \overrightarrow{q}%
_{i}-\overrightarrow{q}_{j}\right| =\sigma $ where $\sigma $ is the
hard-sphere diameter. The relative velocity $\overrightarrow{v}_{ij}^{\prime
}$ after the collision is related to that before the collision by%
\begin{equation}
\overrightarrow{v}_{ij}^{\prime }=\overrightarrow{v}_{ij}-\left( 1+\alpha
\right) \widehat{q}_{ij}\widehat{q}_{ij}\cdot \overrightarrow{v}_{ij}
\label{1}
\end{equation}%
where $\widehat{q}_{ij}=\overrightarrow{q}_{ij}/q_{ij}$ is a unit vector and
the total momentum is unchanged. The parameter $\alpha $ is the coefficient
of (normal)\ restitution: elastic hard spheres correspond to $\alpha =1$
while for smaller values of $\alpha $, energy is lost and the collisions are
inelastic. More complex models allow for velocity-dependent coefficients of
restitution and non-conservation of energy related to the tangential
velocities as well but the focus here is on demonstrating the phenomena in
the simplest model. The simulations were performed in a cubic cell with
walls of length $L$ using Lees-Edwards boundary conditions\cite{LeesEdwards}
which are periodic boundaries in the Lagrangian frame defined by the
velocity field $\overrightarrow{v}\left( \overrightarrow{r}\right) =ay%
\widehat{x}$, where $a$ is the shear rate. It is easy to show that the
balance laws admit of a stationary, spatially homogeneous state with the
collisional cooling balancing the viscous heating to give a constant
temperature. This balance relates the temperature, applied shear rate and
coefficient of restitution so that there are effectively only three
independent dimensionless parameters which will here be taken to be the
number of atoms, $N$, the size of the cell, $L/\sigma $, and the coefficient
of restitution $\alpha $. The temperature can always be set to $k_{B}T=1$ by
adjusting the time scale and the shear rate is then fixed by the energy
balance relation. The procedure in all simulations was to first equilibrate
an equilibrium $(\alpha =1$ and so $a=0$)\ liquid of the desired number of
atoms and cell size by starting with a face-centered cubic configuration and
allowing it to melt over a run of $5\times 10^{7}$ collisions. Long before
the end of these simulations, the system exhibits the expected pressure and
diffusion constant of a hard sphere liquid. The shear rate and coefficient
of restitution are then set to the desired values and the simulations
continued. 

\begin{figure}
\includegraphics[angle=-90,scale=0.3]{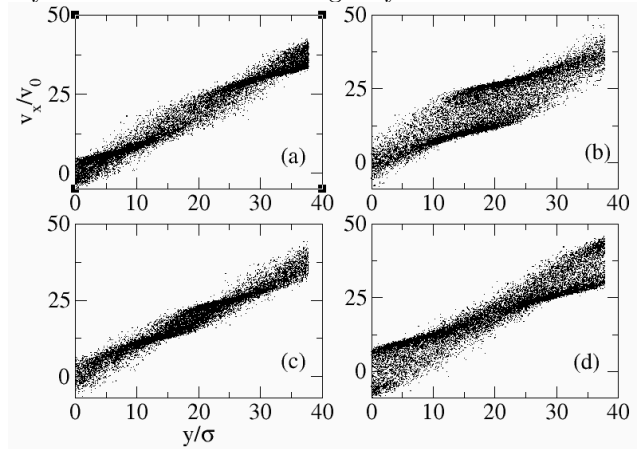}
\caption{\label{fig2} Snapshots of the flow velocity as a function of position along the y
axis. The velocities are scaled to the initial thermal velocity. Each point
is one of the 13500 grains.}
\end{figure}

A simple global measure of the behavior of the system is provided by the
temperature defined as the kinetic energy calculated relative to the assumed
linear shearing profile. Figure 1 shows the temperature as a function of
time for $\alpha =0.7$. The temperature of a small system, made up of 500
grains, is constant whereas that of the larger system, containing 13500
grains, increases rapidly and then appears to oscillate. The large deviation
from between the two is an indication that the assumed profile may be
incorrect for the larger system. When $\alpha $ is set to $0.9$, the larger
system also appears to be stable. Figure 2 shows a snapshot of the velocity
in the $x-$direction as a function of position in the $y$-direction for
systems over one cycle of temperature oscillation while Fig. 2 shows the
peculiar velocity $v_{x}-ay$ as a function of $z$ for the same snapshots.
The $v_{x}$ vs $y$ plot shows a generally linear profile but with clear
signs of local structure which is presumably a manifestation of the kind of
2-D structure described previously\cite{TanGold}. The plot vs. $z$ shows a
clear sinusoidal variation with a time-varying amplitude that grows to many
times the thermal velocity at its maximum. From the density of points in the
figure, it is clear that there is some clustering at the positions of the
maxima and minima of the curves. The snapshots taken at the minimum of the
temperature cycle show that the 2-D structure persists but that the 3-D
structure is virtually absent. Examination of numerous snapshots confirms
that this oscillatory behavior takes place throughout the duration of the
simulation and is the source of the temperature oscillations. As would be
expected from Fig. 1, which shows that the temperature profiles for a small
system of 500 grains at $\alpha =0.7$ and and a 13500 grain system with $%
\alpha =0.9$ are flat, neither of these systems exhibits this structure.
Simulations of the larger system for $\alpha =0.6$ show temperature
variations that are more than 5 times greater than those for $\alpha =0.7$.

\begin{figure}
\includegraphics[angle=-90,scale=0.3]{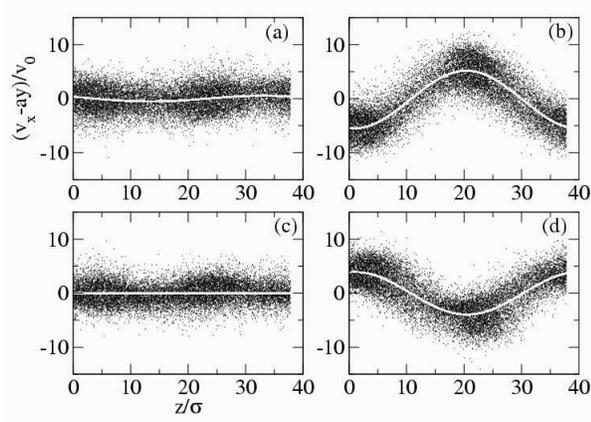}
\caption{\label{fig3} Snapshots of the peculiar velocity in the x-direction as a function
of position along the z axis. The velocities are scaled to the initial
thermal velocity. Each point is one of the 13500 grains. The white line is a
fit to a sin-function with wavevector fixed at $2\pi /L$.}
\end{figure}

To understand why some of the simulations would spontaneously develop the
3-D structure, it is natural to examine the linear stability of the system.
The local hydrodynamic fields, the density $n\left( \overrightarrow{r}%
,t\right) $, temperature $T\left( \overrightarrow{r},t\right) $and the
velocity $\overrightarrow{U}\left( \overrightarrow{r},t\right) $, satisfy
the exact balance equations\cite{GarzoDufty},\cite{LutskoJCP}%
\begin{widetext}
\begin{eqnarray}
\frac{\partial }{\partial t}n+\overrightarrow{\nabla }\cdot \overrightarrow{U%
}n &=&0  \label{2} \\
\frac{\partial }{\partial t}T+\overrightarrow{U}\cdot \overrightarrow{\nabla 
}T+\frac{2}{Dnk_{B}}\overleftrightarrow{P} &:&\overrightarrow{\nabla }%
\overrightarrow{U}+\frac{2}{Dnk_{B}}\overleftrightarrow{\nabla }\cdot 
\overrightarrow{Q}=-\xi   \nonumber \\
\frac{\partial }{\partial t}\overrightarrow{U}+\overrightarrow{U}\cdot 
\overrightarrow{\nabla }\overrightarrow{U}+\frac{1}{mn}\overrightarrow{%
\nabla }\cdot \overleftrightarrow{P} &=&0  \nonumber
\end{eqnarray}%
\end{widetext}where $\overleftrightarrow{P}$ is the pressure tensor, $%
\overrightarrow{Q}$ is the heat flux vector and $\xi <0$ is the cooling
rate. Introducing the peculiar velocity $\overrightarrow{u}\left( 
\overrightarrow{r},t\right) =\overrightarrow{U}\left( \overrightarrow{r}%
,t\right) -\overleftrightarrow{a}\cdot \overrightarrow{r}$ and using the
fluxes and cooling rate derived from the Enskog equation by expanding about
HCS\cite{GarzoDufty} 
\begin{widetext}
\begin{eqnarray}
P_{ij} &=&p\delta _{ij}-\eta \left( a_{ij}+a_{ji}\right) -\eta \left(
\partial _{i}u_{j}+\partial _{j}u_{i}-\frac{2}{3}\delta _{ij}\overrightarrow{%
\nabla }\cdot \overrightarrow{u}\right) -\gamma \delta _{ij}\overrightarrow{%
\nabla }\cdot \overrightarrow{u}  \label{3} \\
Q_{i} &=&-\mu \partial _{i}n-\kappa \partial _{i}T  \nonumber \\
\xi  &=&\xi ^{(0)}+\xi ^{(1)}\overrightarrow{\nabla }\cdot \overrightarrow{u}
\nonumber
\end{eqnarray}%
\end{widetext}where $p$ is the pressure, $\eta $ is the shear viscosity, $%
\gamma $ is the bulk viscosity, $\kappa $ is the thermal conductivity and $%
\mu $ is a transport coefficient which goes to zero at $\alpha =1$ and
describes the fact that density gradients drive heat flow in the granular
system. All of these quantities are functions of the local hydrodynamic
fields. Expanding these fields about steady,uniform values as $n\left( 
\overrightarrow{r},t\right) =n_{0}+\epsilon n_{1}\left( \overrightarrow{r}%
,t\right) +...$, where $\epsilon $ is a fictitious small parameter used to
order a perturbative expansion of Eqs.(\ref{2}-\ref{3}), one finds at order $%
\epsilon ^{0}$ that the only nontrivial constraint is the energy balance
equation%
\begin{equation}
\eta _{0}a^{2}=\frac{3n_{0}}{2}\xi _{0}  \label{4}
\end{equation}%
where $\eta _{0}$ is the shear viscosity evaluated for the zeroth-order
fields. This gives the constraint relating the shear rate to the cooling
rate, and hence to the temperature and the coefficient of restitution. Note
that this uniform solution is consistent with the periodic boundary
conditions in the co-moving frame used in the simulations. Further details
of the calculation, which is straightforward, will be given elsewhere and
here, only the main steps and results summarized. At linear order, it is
convenient to switch to a Fourier representation but the stability analysis
remains nontrivial due to the linear mode-coupling which occurs via the
gradient term arising from the convective terms in eq.(\ref{3}). However, to
investigate stability in the $z-$direction, we can begin by restricting
attention to wavevectors in that direction $\overrightarrow{k}=k\widehat{z}$
in which case, several simplifications occur. First, the gradient term does
not contribute, so that temporal stability is simply determined by the signs
of the eigenvalues of the propagation matrix, the elements of which are
quadratic functions of the wavevector. Second, the $x-$ and $y-$ components
of the velocity. representing shear modes, decouple form the other variables
and are purely decaying with time constant equal to the kinematic shear
viscosity is $\nu _{0}=\frac{1}{mn_{0}}\eta _{0}$. The remaining three
eigenvalues are the roots of a cubic secular equation with $k$-dependant
coefficients. Setting $k=0$ shows that two modes go to zero and the third is
stable with a finite decay constant. The latter assures that small,
homogeneous violations of the energy balance relation eq.(\ref{4}) decay.
The stability of the modes can be determined by looking for solutions of the
secular equation for which the real part of the eigenvalue, $\lambda $, is
zero, or in other words $\lambda =i\xi $ for real $\xi $. Then, the secular
equation splits into two equations, the real and imaginary terms, which
serve to determine both $\xi $ and the critical wavevector $k_{c}$ for which
the mode does not decay. As expected, there are three possible solutions. In
the first, $\xi =0$ and the critical wavevector $k_{c}$ satisfies%
\begin{widetext}
\begin{equation}
\left[ \frac{\xi _{n}^{(0)}}{\xi _{0}^{(0)}}-\frac{\nu _{n}}{\nu }%
-\allowbreak \frac{p_{n}}{p_{T}}\left( \frac{\xi _{T}^{(0)}}{\xi _{0}^{(0)}}-%
\frac{\nu _{T}}{\nu }\right) \right] \frac{3n_{0}}{2}\xi
_{0}^{(0)}=\allowbreak \left( \frac{p_{n}}{p_{T}}\kappa _{0}-\mu _{0}\right)
k_{c}^{2}  \label{7b}
\end{equation}%
\end{widetext}where the shear rate has been eliminated using the energy
balance equation. Here, a non-numerical subscript indicates a derivative
evaluated in the homogenous state, e.g. $\nu _{n}=\left( \frac{\partial }{%
\partial n}\nu \right) _{n=n_{0}}$. Recall that $\mu _{0}$ vanishes for $%
\alpha =1$ so, for small deviations from elastic hard spheres, the
coefficient on the right will be positive and, since $\xi _{0}^{(0)}>0$, the
critical wavevector exists if and only if the term in brackets is positive.
It is easy to show by differentiating the secular equation that the slope of
the real part of the eigenvalue at the critical wavevector is positive so
that the mode is unstable for wavevectors smaller than the critical
wavevector. The remaining two possibilities for critical wavevectors
correspond to complex-conjugate propagating modes which become unstable but
they are only relevant at extremely small values of $\alpha $ and will not
be considered further here. Evaluation of this expression for the parameters
of the simulation using the transport coefficients of ref.\cite{GarzoDufty}
gives $k_{c}\sigma =0.238$. The instability will not be present for small
systems since the smallest nonzero wavevector that can be sampled is $2\pi
/L $ so that this defines a maximum stable system size or minimum number of
grains in a given geometry. For a cubic system, this gives a stability limit
of $N_{c}=4612$ which is indeed between the sizes of systems examined here.
Similarly, for fixed $N=13500$, the instability exists for $\alpha <0.804$
which is in agreement with the observations above. These calculations also
show that the left hand side is positive for all $\alpha $ at zero density.
Nevertheless, this calculation cannot be taken too seriously as the
dependence of the transport properties on the shear rate has not been taken
into account, although it might be hoped that the various ratios appearing
in eq.(\ref{7b}) would be less sensitive to this omission than the absolute
values of the quantities.

The expression for the critical wavevector given in eq.(\ref{7b}) is closely
related to a similar instability that occurs in the hydrodynamic equations
linearized about the HCS\cite{DeltourHCS},\cite{ErnstBritoHCS},\cite{BreyHCS}%
. Obviously, in the HCS there is no viscous heating to counteract the
cooling of the gas, so the system is not in a steady state and hydrodynamic
modes in the usual sense do not exist. However, the system can be mapped to
a steady state by introducing a new time variable related nonlinearly to the
laboratory time. This introduces source terms into the hydrodynamic
equations which replace the two terms in eq.(\ref{7b}) involving the shear
viscosity but the analysis is otherwise identical. As discussed in detail by
Brey et al\cite{BreyHCS}, this linear instability is not seen in the HCS
because it is always slower to develop than the shear instability. The shear
instability does not exist in sheared granular fluids and so it appears that
the secondary instability identified in HCS is actually manifested.

These results only give information about the stability of the uniform
velocity profile. Once the instability begins to develop, the nonlinear
couplings in the hydrodynamic equations will become important and will
govern the long time behavior. (Burnett terms and other higher order
contributions to the fluxes neglected here will also become important.) In
the present case, it is not clear whether these might exert a stabilizing
influence that counteracts the linear instability, or whether the
oscillations are due to a frustration of the system by the cubic simulation
cell and periodic boundaries. A careful analysis of the effect of the
nonlinear terms following along the lines of those performed for HCS\cite%
{NonlinStabilityHCS},\cite{BreyHCS} is clearly needed.

Very recently, simulations of 3-D shear flows have shown the presence of
density variations along the third dimension (perpendicular to both the flow
and the velocity gradient)\cite{Conway}.The simulations were performed using
hard walls at $y=\pm L/2$ which move in the $\pm x$-direction thus creating
a varying flow velocity between them and are characterized by the several
collision parameters specifying the normal and tangential coefficients of
restitution (governing the amount of energy and angular momentum lost in the
collisions), for both the atom-atom and atom-wall collisions and are also
complicated by the inevitable presence of boundary layers near the walls. It
seems likely that this is a manifestation of the same phenomena.

\bigskip

\begin{acknowledgments}
I would like to thank Jim Dufty and Jean-Pierre Boon for several useful
discussions. This work is supported, in part, by the European Space Agency
under contract number C90105.
\end{acknowledgments}

\bigskip

\bibliographystyle{prsty}
\bibliography{physics}

\end{document}